# Back to mechanisms of superconductivity in low-doped strontium titanate

Lev P. Gor'kov [a]

[a] National High Magnetic Field Laboratory, Florida State University, Tallahassee, FL 32310

We have analyzed two mechanisms proposed recently for superconductivity in doped $SrTiO_3$ - the plasmon-mediated pairing and the potential of an instantaneous attraction between two electrons that owes its origin to the exchange by high-frequency optical phonons. The first approach seems to be self-consistent, but in a limited range of the dielectric constant. The direct instantaneous interaction between two electrons was ascribed to exchange by the band longitudinal (LO) phonons. The supposition was incorrect. The current paper shows that contributions into the pairing matrix element in a form of the net attraction come about only as the result of disorder. The experiment revealed the existence of the mobility edge in doped $SrTiO_3$. Electrons occupying states below the mobility threshold are localized on the lattice defects and thereby are strongly coupled to the lattice vibrations. Emergent localized phonon modes attached to the defect is the inevitable corollary of such electron-lattice coupling. The virtual exchange by quantum of a local mode manifests itself as the effective attraction between two electrons in the conduction band. The total matrix element that is the sum over these processes increases with concentration and can overwhelm the Coulomb repulsion screened by the LO band phonons. That is, the role previously ascribed to the band LO phonons actually belongs to the localized phonon modes.

**Introduction**

Electronic properties of doped $SrTiO_3$ still remain poorly understood. In particular, strontium titanate displays the low-temperature superconductivity at the surprisingly low concentrations of doped electrons [1]. The nature of the physical mechanisms behind the superconductivity in weakly doped $SrTiO_3$ challenged the microscopic theory already for more than half of the century [2].

In this presentation we reconsider the two recently suggested mechanisms of superconductivity in doped strontium titanate: the plasmon- mediated pairing [3] and the pairing that owes its origin to the exchange by the longitudinal (LO) optical phonons [4].

The paper is divided into two parts. We start with the analysis of possible phonon mechanisms. The plasmon mechanism [3] is discussed in the second part.

**I. Interactions between two conduction band electrons mediated by phonons**

*I. a. Various contributions into the matrix element.* Previously, it was speculated that the exchange by several LO phonons in $SrTiO_3$ can renormalize the Coulomb repulsion between the two conduction band electrons into an effective attractive interaction (see Eqs. (3, 6) in [4]). The authors [3] pointed out that for such instantaneous potential [4] to become attractive the contributions into the matrix pairing element from each of LO phonons need to be independent and arbitrary large. We agree with [3] in that for clean $SrTiO_3$ such supposition contradicts to the general theorem according to which the dielectric function of a transparent crystal on the imaginary frequency axis has the positive sign.

The experimental discovery [5] of the mobility edge in doped SrTiO$_3$ gives room for a different solution. At concentrations below the mobility threshold, the doped electrons, instead of filling the conduction band, remain localized. In [4], however, doped SrTiO$_3$ was treated as a homogeneous transparent medium with the new average dielectric constant. Thereby the results [4] indeed contradict to the same general theorem.

Missing in [4] (and in [3], by the by) is the analysis of disorder caused by the 'foreign' dopant atoms. In fact, fluctuations in the concentration of dopants and in their positions from end to end of the sample are inevitable. As the doped charges remain immobile and immersed in the lattice, this gives rise to the appearance of the local phonon modes. The exchange between two electrons from the conduction band by a high frequency local phonon leads to the additional contribution into the matrix element that has the structure of an attractive momentary interaction.

We suggest that the localized phonons take over the role conjectured in [4] to the band LO phonons.

***I.b. Disorder and localized phonons.*** Before proceeding further, it is necessary to stipulate that although the current study is motivated by the peculiarities of superconductivity in doped SrTiO$_3$, below one mostly concern oneself with the self-sufficient issue of superconductivity mediated by phonons localized on some strong defects. The latter point deserves a fuller explanation.

The two interactions competing in the BCS phonon model are the direct Coulomb repulsion between two conduction band electrons and the retarded phonon mediated pairing attraction. In SrTiO$_3$ the Coulomb interaction remains repulsive. However, it has significantly weakened in clean strontium titanate being already screened by the band LO phonons (see [3, 4] and Eqs. (24, 25) below). Electrons doped at a concentration below the mobility edge are all trapped on the lattice defects created by doping and remain immobile. We argue that because of this on the very same defects emerge the localized phonon modes. The exchange by the localized phonons gives rise to the additional independent contribution into the pairing matrix element that obviously has the form of an instantaneous attraction.

The outstanding problem of finding the phonon spectrum of the periodic lattice with a defect, however, still awaits a general solution. The phonon spectrum is known only for few simple models (see, e.g., in [6]). That hinders a rigorous quantitative analysis of such mechanism of superconductivity and makes one to resort instead to the analogy with ordinary superconductors.

***I. c. Electrons and phonons in doped SrTiO$_3$.*** We assume that at low concentrations the defects not overlap. We also assume that attached to the defect is a strong potential for electrons. Doped electrons go into the bound states and localize themselves on a discrete level with a negative energy $-|E_B|$. Being caught in the bound state at a lattice defect, the trapped carriers manifest themselves participating in the localized vibrational modes.

Unlike with the phonons in clean SrTiO$_3$ possessing a momentum, the contributions of localized phonons come independently from different parts of the sample. The frequency of a localized mode lies outside (above) the phonon band of the crystalline SrTiO$_3$.

Having suggested that the role ascribed in [4] to the band LO phonons is actually inherent in localized phonons, our focus is now on the interaction of electrons from the conduction band with localized vibrational modes. The mechanism is studied below at small concentrations of short range defects.

Few, if any is known with certainty regarding the structure and strength of local defects produced, for instance, by a single oxygen vacancy in the oxygen- reduced $SrTiO_{3-y}$ [1,2]. In that follow, the idea of the defect means the lattice imperfection that incorporates changes in the local elastic constants and an attractive potential for electrons strong enough to trap one electron in the bound states.

Recall that in the three-dimensional quantum-mechanical problems the bound states appear only in a strong enough attractive potential. The trapped electrons thereby are tightly coupled with the lattice. Defects are not uniformly distributed and the spectrum of localized vibrational modes in their vicinity can be different in different parts of the sample.

Two electrons from the conduction band can exchange by a localized phonon. Leaping ahead, the virtual exchange by the quantum of such mode can play the significant role in superconductivity of doped $SrTiO_3$. The sign of the corresponding matrix element is negative and the latter bears the character of an instantaneous interaction. Constrains imposed in the perfect crystal [3] on the contributions from the propagating band LO phonons do not apply to the localized modes.

The total matrix element increases with the concentration of donors. There are no other small parameters in the problem. In clean $SrTiO_3$ interaction of the electrons with the band LO phonons has already significantly diminished the direct Coulomb electron-electron repulsion [3, 4]. Therefore, at-not-too-small-concentrations the attraction mediated by the localized phonons prevails as the pairing mechanism in doped $SrTiO_3$.

***I.d. Periodic lattice with a defect.*** In that follows, we usually keep in mind a 3D cubic lattice of structureless atoms with a low concentration of the substitutional atoms as defects.

Let $A(r_n - r_{n'})$ is the matrix of the short range interactions between the atoms in the perfect lattice. Expression for the energy of the elastic deformations is:

$$U_{latt} = \frac{1}{2} \sum \vec{u}(r_n) \cdot \vec{\vec{A}}(r_n - r_{n'}) \cdot \vec{u}(r_{n'}). \quad (1)$$

The summation in (1) runs over the positions of the atoms. The phonon spectrum of the regular lattice is defined by the equation:

$$M\omega^2 \vec{u}(r_n) = \sum_{n'} \vec{\vec{A}}(r_n - r_{n'})\vec{u}(r_{n'}). \quad (2)$$

In the presence of a defect:

$$M\omega^2 \vec{u}(r_n) = \sum_{n'} \left( \hat{A}(r_n - r_{n'})\vec{u}(r_{n'}) + \hat{L}_k(r_n - r_k; r_k - r_{n'})\vec{u}(r_{n'}) \right), \quad (3)$$

where, as above, $\hat{A}(r_n - r_{n'})$ is the matrix of elastic constants for the perfect lattice and $\hat{L}_k(r_n - r_k; r_k - r_{n'})$ stands for the interactions between the defect at the site $r_k$ and the rest of the lattice.

***I.e. Localized modes and their contributions into the pairing matrix element.*** Feasibility of solving Eq. (3) analytically depends on the structure of the elastic matrix for the defect $\hat{L}_k(r_n - r_k; r_k - r_{n'})$. The exact solutions are known only for few simple models (see e.g. [6]). The lack of the general methods, as mentioned above, prevents the actual calculation of the pairing matrix element for superconductivity mediated by localized phonons. There are, however, several results that follow simply from the analogy with the ordinary metals.

The Hamiltonian for an electron moving in the potential of the lattice is:

$$\hat{H} = -\frac{\Delta}{2m_e} + \sum_{i \neq j} U(\vec{r} - \vec{r}_i) + \sum_j V(\vec{r}_j; \vec{r} - \vec{r}_j), \quad (4)$$

where $U(\vec{r} - \vec{r}_i)$ stands for the potential of the atoms of the perfect lattice, and $V(\vec{r}_j; \vec{r} - \vec{r}_j)$ is the potential of a defect ($\vec{r}_j$ -the position of the substitutional atom in the lattice). Electrons scatter on defects and on the lattice vibrations.

The electron-phonon Hamiltonian has the form:

$$\hat{H}_{e-ph}(\vec{r}) = \sum_{i \neq j} \left( \vec{d}U(\vec{r} - \vec{r}_i) / d\vec{u}(\vec{r}_i) \right) \hat{\vec{u}}(\vec{r}_i) + \sum_j \left( \vec{d}V(\vec{r}_s; \vec{r} - \vec{r}_j) / d\vec{u}(\vec{r}_j) \right) \cdot \hat{\vec{u}}(\vec{r}_{ji}). \quad (5)$$

Introduce the notation

$$\hat{M}(\vec{q} \mid \omega(\vec{q})) = (1/\Omega) \int \exp(i\vec{q} \cdot \vec{r}) \hat{H}_{e-ph}(\vec{r}) d^3\vec{r}. \quad (6)$$

$\hat{M}(\vec{q} \mid \omega(\vec{q}))$ denotes the matrix element for scattering of an electron with the momentum $\vec{k}$ on the potential (5) with the transfer of the momentum $\vec{q} = \vec{p} - \vec{k}$ accompanied by emitting or absorprtion one phonon with a frequency $\omega(\vec{q})$. The wave functions of electrons are normalized on the volume $\Omega$ of the system.

Rewriting (5), one obtains after simple transformations:

$$\hat{M}(\vec{q} \mid \omega(\vec{q})) = (1/N) \left\{ \sum_i \exp(i\vec{q} \cdot \vec{r}_i)(\vec{Z} \cdot \hat{\vec{u}}(\vec{r}_i)) + \sum_j \exp(i\vec{q} \cdot \vec{r}_j)(\vec{K} \cdot \hat{\vec{u}}(\vec{r}_j)) \right\}. \quad (7)$$

In Eq. (7)
$$\vec{Z} = \int \exp(i\vec{q} \cdot \vec{r}) \left[ \frac{\vec{d}U(\vec{r})}{d\vec{r}} \right] \frac{d^3\vec{r}}{\Omega_0} \quad (8a)$$

and

$$\vec{K} = \int \exp(i\vec{q} \cdot \vec{r}) \left[ \frac{dV(\vec{r})}{d\vec{r}} \right] \frac{d^3\vec{r}}{\Omega_0} \qquad (8b)$$

are the two constant vectors characterizing the short range potentials of atoms of the regular lattice and, correspondingly, that of the defect; $\Omega_0$ is the volume of the unit cell.

For the perfect lattice the expression of the displacements' operators in the second quantization is [7]:

$$\hat{\vec{u}}^\alpha(\vec{r}_n) = \frac{1}{N^{1/2}} \sum_k \frac{1}{\sqrt{2M\omega(\vec{k})}} \{\hat{b}_{k,\alpha} \vec{e}^{(\alpha)}(\vec{k}) \exp(i\vec{k} \cdot \vec{r}_n) + \hat{b}^+_{k,\alpha} \vec{e}^{(\alpha)*}(\vec{k}) \exp(-i\vec{k} \cdot \vec{r}_n) \}, \quad (9)$$

where $\hat{\vec{u}}^\alpha(\vec{r}_n)$ is one of the three vector components of $\hat{\vec{u}}(\vec{r}_n)$, $\omega(\vec{k})$ is the phonon frequency. Substituting (9) into Eq. (7) one obtains:

$$\hat{M}(\vec{q} \mid \omega(\vec{q})) = \frac{1}{N^{1/2} \sqrt{2M\omega(\vec{q})}} (\vec{Z} \cdot \vec{e}(\vec{q})). \qquad (10)$$

The phonon-mediated pairing matrix element is of the second order in $\hat{M}(\vec{q} \mid \omega(\vec{q}))$:

$$W(\vec{q} \mid \omega(\vec{q})) = |\hat{M}(\vec{q} \mid \omega(\vec{q}))|^2 \times \left[ \frac{1}{E_0 - E_I} + \frac{1}{E_0 - E_{II}} \right]. \qquad (11)$$

Here $E_0 = \varepsilon_1(\vec{k}) + \varepsilon_2(-\vec{k})$ is the energy of the initial state of two electrons; $E_I = \varepsilon_1(\vec{k}) + \omega(\vec{q})$ and $E_{II} = \varepsilon_2(-\vec{k}) + \omega(\vec{q})$ are energies of the two intermediate states. After substitution of Eq. (10) one finds:

$$W(\vec{q} \mid \omega(\vec{q})) = \frac{1}{N} \left[ \frac{(\vec{Z} \cdot \vec{e}(\vec{q}))}{\sqrt{M}} \right]^2 \times \left[ \frac{1}{(\varepsilon_1(\vec{k}))^2 - \omega(\vec{q})^2} \right]. \qquad (12)$$

At low temperatures when $(\varepsilon_1(\vec{k}))^2, (\varepsilon_1(\vec{k}-\vec{q}))^2 \ll \omega(\vec{q})^2$ one arrives to the expression:

$$W(\vec{q} \mid \omega(\vec{q})) \approx -\frac{1}{N\omega(\vec{q})^2} \left[ \frac{(\vec{Z} \cdot \vec{e}(\vec{q}))}{\sqrt{M}} \right]^2 < 0. \qquad (13)$$

The parameter $\lambda$ in the canonic weak coupling expression for temperature of the superconducting transition:

$$T_C = \varpi \exp(-1/\lambda) \qquad (14)$$

equals

$$\lambda = \frac{\Omega_0 Z^2 m_e p_F}{2\pi^2 M \omega_0^2} \quad (15)$$

The dispersion in the phonon frequency was omitted $\omega(\vec{q}) \simeq \omega_0$; the prefactor $\varpi$ in Eq.(14) is of the order of $\omega_0$. Usually, the parameter $\lambda$ is not small.

Note the dependence on $N$ in the denominators in the expressions (10, 13): in Eq. (10) the factor $N^{-1/2}$ comes about from the fact that both the electron wave functions and the plane wave in (9) extend over the whole space. The conservation law gives the factor $N^{-1}$ in Eq. (13).

In the presence of a defect one would expect, instead of the plane waves in Eq. (9), the new expression for the displacements in a form:

$$\hat{\vec{u}}(\vec{r}_n) = \frac{1}{N^{1/2}} \sum_s \frac{1}{\sqrt{2M\omega_s}} \left\{ \hat{b}_{s,\alpha} \vec{e}_s^{(\alpha)}(\vec{k}) A_s(\vec{r}_n) + \hat{b}_{s,\alpha}^+ [\vec{e}_s^{(\alpha)}]^* A_s^*(\vec{r}_n) \right\}, \quad (16)$$

where $A_s(\vec{r}_n)$, $A_s^*(\vec{r}_n)$ are the new eigenfunctions corresponding to a frequency $\omega_s$. As far as $\omega_s$ belong to the continuous phonon band spectrum, the matrix element $\hat{M}(\vec{q}|\omega_s)$ in Eq. (6) retains the factor $N^{-1/2}$ in front (and the proportionality to $1/N$ in $W(\vec{q}|\omega_s)$) because the significant changes in the pattern of displacements take place only in a vicinity of the defect. For the localized mode $\omega_s^{loc}$ $A_s^{loc}(\vec{r}_n)$ decreases away from the defect position. Correspondingly, in that case $\hat{M}(\vec{q}|\omega_s) \propto 1/N$ and $W(\vec{q}|\omega_s) \propto 1/N^2$.

Contributions into the total pairing matrix element from each part of the sample are independent. It then follows that in an order of magnitude:

$$W^{tot}(\vec{q}|\omega_s) \approx c W(\vec{q}|\omega(\vec{q})), \quad (17)$$

where $c$ is the concentration of defects.

*I.f. Implications for doped SrTiO$_3$.* Return to superconductivity in strontium titanate. The dimensionless parameter $\lambda$ in the BCS-like Eq. (14) for $T_C$ is the product of the electron-phonon scattering amplitude and the density of states at the Fermi surface. In low doped SrTiO$_3$ the former is of the atomic scale, as in most metals, but the Fermi surface is small [8]. At first glance the long-range interaction with a longitudinal LO optical mode in the polar SrTiO$_3$ can compensate smallness of density of states on the Fermi surface, as it was first suggested in [9] and discussed in [4]. However, the long range interactions do no more than just screen the direct Coulomb repulsion in clean SrTiO$_3$. The interactions of the electrons with band LO phonons generate no effective net attraction [3].

Keeping in mind SrTiO$_3$, let us stay longer on possible localized phonons in a polar crystal. At the choice of the lattice model for this problem one is to remember that both the polarizable lattice atoms and the polarizable defects possess the internal degrees of freedom. To the best of the author's knowledge, no attempts have been made to address the issue of the phonon spectrum of a defect in the lattice with several atoms per unit cell. Some of the qualitative results can be outlined as in [10].

Limit ourselves by a lattice with two atoms per unit cell and let the polarization be proportional to the relative displacement $\vec{u}_s(\vec{r}_n) = \vec{u}_1 - \vec{u}_2$. Following [10], consider three contributions into the energy of the lattice. The first one is the energy of the elastic deformations. The exact form of the elastic matrix in the new Eqs. (1-3) is not important. It is postulated that the localized mode came about already from the solution of the problem with short range elastic interactions. The vector $\vec{r}_n$ in the notation $\vec{u}_s(\vec{r}_n)$ signifies the center of the unit cell.

The polarizability of the lattice gives rise to the appearance of the macroscopic electrical fields $\vec{E}_s(\vec{r})$ and, in turn, to the new terms in the lattice energy [10]:

$$U = \sum_n q_e (\vec{u}_s(\vec{r}_n) \cdot \vec{E}_s(\vec{r}_n)) - \frac{1}{8\pi}(\kappa_\infty - 1)\int (\vec{E}_s(\vec{r}))^2 d^3\vec{r} \ . \quad (18)$$

(Here $(\kappa_\infty - 1)/4\pi$ is the polarizability at the fixed position of the atoms). In (18) $q_e$ is a charge and $\vec{P}(\vec{r}_n) = q_e \vec{u}_s(\vec{r}_n)$ is the phenomenological expression for the polarization that is specific to a given material.

For simplicity, consider $\vec{u}_s(\vec{r}_n)$ in (18) as a continuous function of the coordinate $\vec{r}$. Rewriting (18) obtains:

$$U = \int \rho_e(\vec{u}_s(\vec{r}) \cdot \vec{E}_s(\vec{r})) d^3\vec{r} - \frac{1}{8\pi}(\kappa_\infty - 1)\int (\vec{E}_s(\vec{r}))^2 d^3\vec{r} \quad (18a)$$

($\rho_e$ in (18a) is the charge density).

The potential energy of a conduction band electron in the presence of the electric fields in the lattice is $-e\varphi(\vec{r})$, where $\varphi(\vec{r})$ is the electric field potential $\vec{E}(\vec{r}) = -\vec{\nabla}\varphi(\vec{r})$.

The longitudinal electric fields obey the equation:

$$div\vec{D}(\vec{r}) = 0. \quad (19)$$

$\vec{D}(\vec{r})$ is:

$$\vec{D}(\vec{r}) = \vec{E}(\vec{r}) + 4\pi \vec{P}(\vec{r}). \quad (20)$$

For the polarization $\vec{P}(\vec{r}) = -\delta U / \delta \vec{E}(\vec{r})$ one finds from Eq. (18a):

$$\vec{P}_s(\vec{r}) = \rho_e \vec{u}_s(\vec{r}) + (\kappa_\infty - 1)\vec{E}_s(\vec{r})/4\pi. \quad (21)$$

Substitution of (21) into Eqs. (19, 20) gives the equation defining the potential $\varphi(\vec{r})$:

$$\kappa_\infty \Delta \varphi(\vec{r}) = 4\pi \rho_e (div \vec{u}_s(\vec{r})). \quad (22)$$

Note that there is the qualitative difference between the matrix elements of the electron-phonon interaction for the band and localized phonons. For the plane wave $\vec{P}_s(\vec{r}) \propto \exp(i\vec{q} \cdot \vec{r})$:

$$\hat{M}(\vec{q} \mid \omega(\vec{q})) = -ieN^{-1/2}\vec{P}_s(q)/q \propto e/q. \quad (23)$$

In the case of a localized mode the polarization is non-zero only in a vicinity $a$ of the defect. At $qa \gg 1$ $\hat{M}(\vec{q} \mid \omega(\vec{q})) \propto e/q$ as in (23), and $\hat{M}(\vec{q} \mid \omega(\vec{q})) \propto ea$ at $qa \ll 1$.

## II. Plasmon mechanism

***II.a. An alternative analysis of the plasmon mediated pairing.*** The broadband insulator strontium titanate possesses a number of unique properties. Among them is the very large static dielectric constant that at helium temperatures reaches 25000 [11]. There are three LO optical modes in the cubic SrTiO$_3$. In that follows the most important is the interaction of electrons with the high frequency LO mode $\omega_{LO}$ of the order of $100 meV$ [12, 13].

The temperature $T_C$ of the superconductivity transition in doped SrTiO$_3$ was calculated in the frameworks of the plasmon-mediated pairing concept. The physical idea underlying the model [3] and some difficulties with which it meets at applications can as we demonstrate, be understood without resorting to the numeric calculations. As far as the static dielectric constant is large [11], the plasmon mechanism [3] reveals no inconsistencies. In practice, in doped SrTiO$_3$ the dielectric constant is not so large. Additional assumptions were necessary in [3] to reproduce the experimental dependence of $T_C$ on doping. Therefore the numeric calculations [3], in our opinion, do not contribute to the concept on a qualitative level and mainly serve to fit the data. In short, we find the plasmon mechanism self-consistent in a narrow range of the value of the dielectric constant.

***II. b. Deriving the plasmon model [3].*** Consider the matrix element of the direct Coulomb interaction between two electrons screened via the exchange by a LO phonon with the frequency $\omega_{LO}$ [9]:

$$\Gamma_s(q, \omega_{nm}) = \frac{4\pi e^2}{\kappa_\infty q^2} - \frac{4\pi e^2}{q^2}\left(\frac{1}{\kappa_\infty} - \frac{1}{\kappa_0}\right) \times \frac{\omega_{LO}^2}{\omega_{LO}^2 + (\omega_n - \omega_m)^2}. \quad (24)$$

At low temperatures $\omega_{nm}^2 \ll \omega_{LO}^2$ and $\kappa_0 \simeq 25000$ [11] the sum is the weak repulsive Coulomb potential:

$$\Gamma_s(p \mid k) = \frac{4\pi e^2}{\kappa_0 q^2} > 0. \quad (25)$$

($|q| = |\vec{p} - \vec{k}|$). Screening of $\Gamma_s(p \mid k)$ Eq. (25) by the accumulating charges is treated in the Random Phase Approximation (RPA):

$$\tilde{\Gamma}_{scr}(q,\omega_{mn}) = \frac{4\pi e^2}{Q^2(q,\omega_{mn})\kappa_0} \quad (26)$$

In the denominator of (26) $Q^2(q,\omega_{mn}) = q^2 + \kappa_{TF}^2 S(q,\omega_{mn})$; $S(q,\omega_{mn})$ is the electronic polarization:

$$S(q,\omega_{mn}) = \int_0^1 \left[(\vec{v}_F \cdot (\vec{p}-\vec{k}|))^2 / [(\vec{v}_F \cdot (\vec{p}-\vec{k}))^2 + \omega_{mn}^2]\right] d\mu,\ \kappa_{TF}^2\text{ is the square of the Thomas-Fermi}$$

radius $\kappa_{TF}^2 = (4e^2 m p_F / \bar{\kappa}_0 \pi \hbar^3)$.

Introduce the effective Bohr radius $\bar{a}_B = \kappa_0 \hbar^2 / e^2 m$ and define the dimensionless parameter $x = (\pi \bar{a}_B p_F / \hbar)$; one has $\kappa_{TF}^2 = 4 p_F^2 / x$. With the dielectric constant $\kappa_0$ of clean SrTiO$_3$ equal 25000 the hydrogenic Bohr radius $\bar{a}_B = \kappa_0 (m_e / m) a_B$ is large $\bar{a}_B \approx 0.7 \times 10^{-4} cm$ ([5]). At concentrations of the carries in [1, 10] the parameter $x$ is very large, too $x \gg 1$.

At small $(v_F q)^2 \ll \omega_{mn}^2$ one has $Q^2(q,\omega_{mn}) = q^2 + \kappa_{TF}^2 (v_F^2 q^2 / 3\omega_{nm}^2)$. Denote $\tilde{\omega}_{pl}^2 = \kappa_{TF}^2 v_F^2 / 3$ as the square of the "plasmon" frequency. After all substitutions, one finds:

$$\tilde{\omega}_{pl} = (8/3x)^{1/2} \varepsilon_F \approx (1.63 / x^{1/2}) \varepsilon_F. \quad (27)$$

Then $Q^2(q,\omega_{mn}) = q^2(1 + \tilde{\omega}_{pl}^2 / \omega_{nm}^2)$ and the interaction in Eq. (26) acquires the form [3]:

$$\tilde{\Gamma}_{scr}(q,\omega_{mn}) = \frac{4\pi e^2 \omega_{nm}^2}{q^2(\omega_{nm}^2 + \tilde{\omega}_{pl}^2)\kappa_0} \equiv \frac{4\pi e^2}{q^2 \kappa_0}\left[1 - \frac{\tilde{\omega}_{pl}^2}{\omega_{nm}^2 + \tilde{\omega}_{pl}^2}\right] \quad (28)$$

At $x \gg 1$ the plasmon frequency (27) is small compared to the Fermi energy and Eq. (28) allows the BCS-like interpretation as the sum of a "retarded" (that is, the plasmon-mediated) attraction and the repulsive Coulomb potentials. From the theory point of view, Eq. (28) is the exact analog to the phonon-mediated retarded interaction in ordinary superconductors [3].

*II.c. Weak coupling approach.* The Migdal-Eliashberg equations were solved in [3] numerically. However, provided that at small enough concentrations $\kappa_0$ in the expression $(4\pi e^2 / q^2 \kappa_0)$ in front of Eq. (4) is the same as in clean SrTiO$_3$, it is not unreasonable to view Eq. (28) as the sum of two weak interactions. The textbook expression for the temperature of the superconductivity transition in the weak-coupling approximation is:

$$T_C = 1.14 \bar{\omega}_{pl} \exp(-1/\lambda) \equiv 1.14(1.63/x^{1/2})\varepsilon_F \exp(-1/\lambda) \quad (29)$$

In the logarithmic approximation $\lambda$ can obtained directly. In the equation for the gap function

$$\Psi(p) = -T \sum_m \int \tilde{\Gamma}(p|k) G(k) G(-k) \Psi(k) d\vec{k} / (2\pi)^3 \quad (30)$$

the product of the two Green functions is $G(k)G(-k) = 1/v_m^2 + [(\vec{k}^2 - p_F^2)/2m]^2$. The Cooper instability is related to the logarithmic singularity at the summation and integration over $v_m$ and $\varsigma = (\vec{k}^2 - p_F^2)/2m \approx v_F(p - p_F)$. Make the following transformations in r.h.s. of Eq. (30)

$$-T \sum_{n'} \int \tilde{\Gamma}(p|k) \frac{m p_F d\mu d\varsigma}{(2\pi)^2} \times \frac{1}{v_m^2 + \varsigma^2} \Rightarrow \left[ \int_0^\pi \frac{m p_F d\mu}{\pi^2} \tilde{\Gamma}(p|k)|_{FS} \right] \times \int_0^W \frac{d\varsigma}{\varsigma} th \frac{\varsigma}{2T} \Rightarrow \lambda \ln\left( \frac{2W\gamma}{\pi T} \right) \quad . \quad (31)$$

($\mu$ is $\cos\theta$ of the angle between the two vectors $\vec{p}$ and $\vec{k}$; both $\vec{p}$ and $\vec{k}$ are on the Fermi surface). After substitution of the vertex $\tilde{\Gamma}(p|k) \equiv (4\pi e^2 /(\vec{p}-\vec{k})^2 \kappa_0)|_{FS}$ and integration over $\mu$ one obtains

$$\lambda = \frac{1}{x} \ln(1+x) << 1 \quad (32)$$

(Following [3], in the integral over $\cos\theta$ was introduced the cutoff: $(\vec{p}-\vec{k})^2 \to (\vec{p}-\vec{k})^2 + \kappa_{TF}^2$).

The Fermi energy $\varepsilon_F = p_F^2/2m$ can be rewritten with the use of the parameter $x = (\pi \bar{a}_B p_F / \hbar)$ and Eq. (29) acquires the form:

$$T_C = [1.86(\hbar/\pi\bar{a}_B)^2 (1/2m)] \times f_1(x) . \quad (33)$$

The function $f_1(x) = x^{3/2} \exp[-x/\ln(1+x)]$ is shown in Fig. 1. The expressions (27, 29, 32, 33) apply at $x >> 1$.

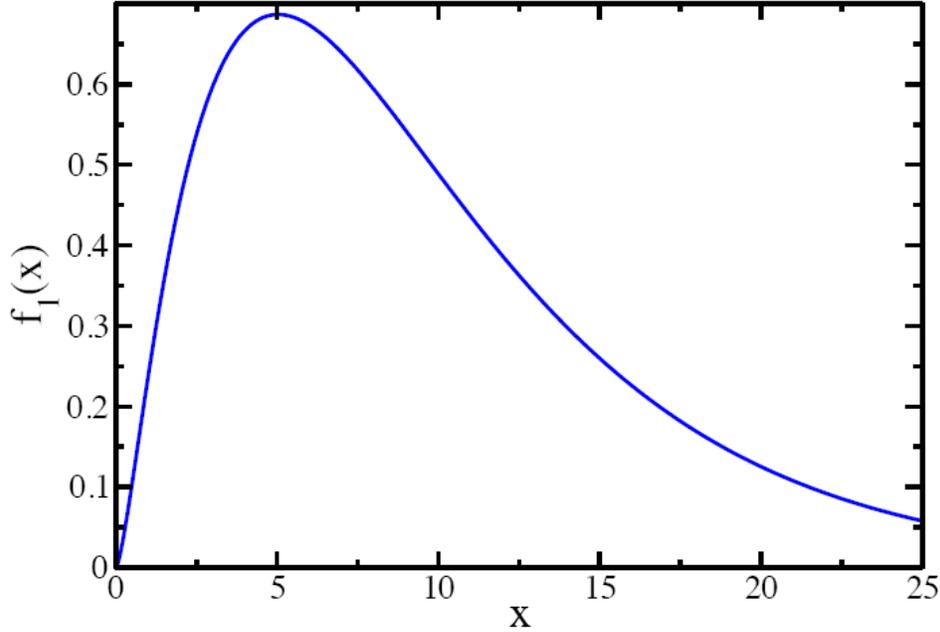

FIG.1. Temperature of the superconductivity transition $T_C$ in one-band model as function of the concentration of electrons. $T_C$ is proportional to $f_1(x) = x^{3/2} \exp[-x/\ln(1+x)]$. The variable $x$ is related to concentration $x = \pi p_F \bar{a}_B / \hbar \simeq \pi^2 \bar{a}_B n^{1/3}$ ; the maximum $f_1(x)$ is at $x \simeq 5$

***II.d. Relevance for the experimental data.*** Compare (27, 29, 32, 33) and the experimental data on the concentration $T_c$ dependence [1, 10].

Recall that at the derivation of the plasmon model Eq. (28) the static dielectric constant was assumed equal to that of clean strontium titanate $\kappa_0 = 25000$. Substituting the hydrogenic Bohr radius $\bar{a}_B \approx 0.7 \times 10^{-4} cm$ into the pre-factor $1.86(\hbar/\pi\bar{a}_B)^2(1/2m)$ in Eq. (33) ($m \approx 1.85 m_e$ [10]) and calculating, one finds:

$T_C \approx 0.75 \times 10^{-4} f_1(x) K.$ (34).

The value of $T_c \sim 10^{-4} K$ from Eq. (33) is by two - three orders smaller in magnitude than observed experimentally [1, 10].

One can approach the problem from the other end, however. Note that experimentally $T_c$ has the maximum at $0.2 K$ at $n \approx 2 \times 10^{18} cm^{-3}$ [8]. Maximum in Fig. 1 for $f_1(x)$ is at $x \simeq 5$. Defining anew the Bohr radius $a_B^{exp}$ from the position of the maximum $x_{max} \simeq \pi^2 a_B^{exp} n^{1/3} = 5$, one finds $a^{exp}_B = 0.5 \times 10^{-6} cm$, much shorter than $\bar{a}_B \approx 0.7 \times 10^{-4} cm$. This time the substitution of $a_B^{exp} = 0.5 \times 10^{-6} cm$ into Eq. (33) leads to a more reasonable value for $T_C \approx 1K$.

From the above estimates, it becomes clear that the static dielectric constant $\kappa_0$ decreases with doping. The authors [3] have arrived to the same conclusion by solving the full set of the Eliashberg equations numerically. For their results to agree with the experimental data at the lowest concentration, the static dielectric constant $\kappa_0$ is to be $\kappa_0 \approx 1000$ or less. With a smaller $\kappa_0$ (i. e., shorter $a_B^{exp}$) increases the Migdal adiabatic parameter $\bar{\omega}/\varepsilon_F$. At $x \simeq 5$ $\bar{\omega}/\varepsilon_F$ in Eq. (27) is of the order of unity and the applicability of the Migdal-Eliashberg equations is violated.

**Discussion**

**A.** The key idea [3] is that at a large enough static dielectric constant the plasmon frequency $\tilde{\omega}_{pl}$ is small compared to the Fermi energy. That is, at calculating $T_C$ one can use the Migdal-Eliashberg machinery as in the ordinary superconductors.

The analysis above has shown that, in practice, $\kappa_0$ decreases with doping. With decreasing $\kappa_0$ increases the Migdal adiabatic parameter $\tilde{\omega}_{pl}/\varepsilon_F$ violating the applicability of the Migdal-Eliashberg equations. According to Eq. (27), $\tilde{\omega}_{pl} = (1.63/x^{1/2})\varepsilon_F$ where in the parameter $x = (\pi \bar{a}_B p_F / \hbar)$ $\bar{a}_B = \kappa_0 \hbar^2 / e^2 m$ is the effective Bohr radius and $p_F \propto n^{1/3}$.

One may expect that the idea would work at higher concentrations or, more generally, at $x \gg 1$. Unfortunately, although the Migdal parameter $\tilde{\omega}_{pl}/\varepsilon_F$ decreases with $x$, $T_C$ decreases exponentially as $T_C \propto \exp(-x/\ln(1+x))$. To summarize, the BCS-like interpretation of Eq. (28) as the sum of a "retarded" plasmon-mediated attraction and the repulsive Coulomb potential is limited to some narrow range of the static dielectric constant $\kappa_0$.

**B.** Returning to the mechanism of localized phonons discussed in the first part of the paper, let us retrace how doping manifest itself in $SrTiO_3$ at different concentrations.

Starting from the smallest concentrations one would expect the Mott insulator-to-metal transition to occur already at low doping $n_s^{MT} < 10^{10} cm^{-3}$, in the accordance with the Mott criterion $n_s^{1/3} a_B^* > 0.26$. The notion of the Mott's impurity band, however, is hardly meaningful in $SrTiO_3$ with its large static dielectric constant. Experimentally, the crossover to conductivity of the coherent band carriers is seen at $n_s^* \approx 2 \times 10^{16} cm^{-3}$ [5]; $n_s^*$ may be taken as an estimate for the mobility edge.

Shubnikov-de Haas (SdH) quantum oscillations (QO) are first seen at $n_s \approx 4 \times 10^{17} cm^{-3}$ [8]. The concentration band is established and the chemical potential enters the conduction band.

QO *and* superconductivity are observable at somewhat higher carrier concentration $n_s \approx 5.5 \times 10^{17} cm^{-3}$. The Fermi energy at such concentrations is $E_F \approx 1.1 \div 1.3 meV$ [1, 8]. According to our interpretation, at these concentrations the attraction mediated by the localized phonons has started to prevail as the pairing mechanism in doped SrTiO$_3$.

$T_C$ increases when $n_s$ varies from $n_s \approx 5.5 \times 10^{17} cm^{-3}$ to $n_s \approx 1.05 \times 10^{18} cm^{-3}$ [8]. After reaching a maximum at $n_{s\max} \approx 2 \times 10^{18} cm^{-3}$ $T_C$ decreases before to grow again as if the conduction electrons begin to fill the next band. Observation of the second SdH frequency at $n_{c1} \simeq 1.2 \times 10^{18} cm^{-3}$ confirms that the chemical potential touches the bottom of the second band. Another QO frequency above $n_{s2} \approx 2 \times 10^{19} cm^{-3}$ signals that the chemical potential has reached the bottom of the third band [8].

Maxima in the $T_c$-dependence is most remarkable feature among the experimental findings [1,8]. Assume that there is a sharp boundary at the mobility edge between the localized and delocalized electronic states. With the manifold of all localized electronic states being occupied, the spectrum of the localized phonon modes is not changing at the further doping. $T_c$ initially increases with the number of carriers in the conduction band, and decreases when accumulating charges screen the interactions between them. This naive mechanism , however, suggests the natural interpretation of maxima in the concentration dependence of $T_c$.

In short, the supposition regarding the leading role of localized phonons in the mechanisms of superconductivity in doped SrTiO$_3$ is not contradictory to the experimental data. The exchange by the phonon modes localized on a three dimensional defect gives a contribution into the instantaneous attractive pairing potential. If this contribution into the total matrix element exceeds that of the screened Coulomb scattering, the superconductivity is due to this mechanism.

## Acknowledgments

The work was supported by the National High Magnetic Field Laboratory through NSF Grant No. DMR-1157490, and the State of Florida.